\documentclass[12pt]{article}
\usepackage{graphicx}

\begin{document}
\centerline{\bf Ising model with spins $S=1/2$ and $1$}
\centerline{\bf on directed and undirected Erd\"os-R\'enyi random graphs}

\bigskip
\centerline{F.W.S. Lima$^1$ and M.A.Sumour$^2$,}

\bigskip
\noindent
$^1$ Dietrich Stauffer Computational Physics Lab,
Universidade Federal do Piau\'{\i}, 64049-550, Teresina - PI, Brazil \\
$^2$ Physics Department, Al-Aqsa University
P. O. Box 4051, Gaza, Gaza Strip, Palestinian Authority\\
\medskip
  e-mail:  fwslima@gmail.com, msumoor@alaqsa.edu.ps
\bigskip

{\small Abstract: Using Monte Carlo simulations we study the Ising model with 
spin $S=1/2$ and $1$ on {\it directed} and {\it undirected} Erd\"os-R\'enyi (ER)
random graphs, with $z$ neighbors for each spin. 
In the case with spin $S=1/2$, the {\it undirected} and {\it directed} ER graphs present a spontaneous magnetization in the universality class of mean field theory, where in both {\it directed} and {\it undirected} ER graphs the model presents a spontaneous magnetization at $p = z/N$ ($z=2, 3, ...,N$), but no spontaneous magnetization at $p = 1/N$ which is the percolation threshold.
For both {\it directed} and {\it undirected} ER graphs with spin $S=1$ we find a 
first-order phase transition for $z=4$ and $9$ neighbors.}


\bigskip

{\bf Introduction}

This paper deals with Ising spins on both directed and undirected Erd\"os-R\'enyi
(ER) graphs. Sumour and Shabat \cite{sumour,sumourss} investigated Ising
models with spin $S=1/2$ on {\it directed} BA networks \cite{ba} using the usual Glauber
dynamics. No spontaneous magnetization was found, in contrast to the case of {\it
undirected} BA networks \cite{alex,indekeu,bianconi} where a spontaneous magnetization was
found below a critical temperature which increases logarithmically with the system size.
For $S=1/2$ systems on {\it undirected} Small-World networks (SW) \cite{nihat} with scale-free
hierarchical-lattice, conventional and algebraic (Berezinskii-Kosterlitz-Thouless)
ordering, with finite transition temperatures, have been found. Lima and Stauffer
\cite{lima} simulated {\it directed} square, cubic and hypercubic lattices ranging from
two to five dimensions with heat bath dynamics in order to separate the network effects from directedness. They also compared different spin-flip algorithms, including cluster
flips, for Ising-BA networks. They found a freezing-in of the magnetization
similar to  the one in Ref.\cite{sumour,sumourss}, following an Arrhenius law at least in
low dimensions. This lack of a spontaneous magnetization (in the usual sense) is
consistent with the fact that if on a directed lattice a spin $S_j$ influences spin
$S_i$, then spin $S_i$ in turn does not influence $S_j$, and there may be no well-defined
total energy. Thus, they showed that for the same  scale-free networks, different
algorithms give different results. Lima et al. \cite{lima2} studied the Ising
model for spin $S=1$, 3/2 and 2 on {\it directed} BA network. The Ising model with spin
1, 3/2 and 2 seemed not to show  a  spontaneous magnetization and their decay time
for flipping of the magnetization followed an  Arrhenius law for heat bath algorithms
that agrees with the results of the Ising model for spin $S=1/2$ \cite{sumour,sumourss} on
directed BA network. S\'anchez et al. \cite{sanches} on {\it directed} SW obtained a 
second-order phase transition for values of rewiring probability  $p=0.1$ and a first-order phase   transition for $p=0.9$ with $p_{c} \approx 0.65$ for the change of phases. The magnetic properties of Ising models defined on the triangular Apollonian network was
investigated for Andrade and Herrmann \cite{her} and no evidence of phase transition was found. In this work, we have studied the Ising model with spins $S=1/2$
and $1$ on  directed and undirected
ER graphs. {\it Undirected} ER graphs  with spin $S=1/2$ present a spontaneous magnetization in the universality class of mean field theory and for $S=1$, we find evidences of first-order phase transition for $z\ge 2$. {\it Directed} ER graphs for spin $S=1/2$ and $S=1$ present a spontaneous magnetization for $z\ge 2$.
Here $z$ is the number of neighbors for each spin.

\begin{figure}[hbt]
\begin{center}
\includegraphics [angle=-90,scale=0.5]{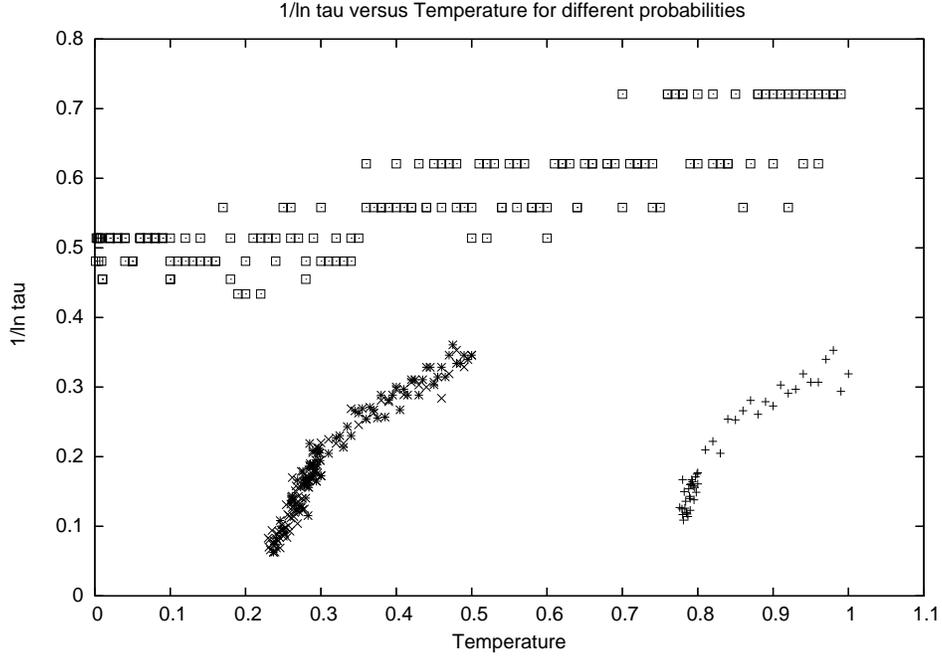}
\end{center}
\caption{$1/\ln (\tau)$ versus temperature for different probabilities $1/N$ (sq.), $2/N$ (x), and $3/N$ (+); directed ER with $S=1/2$.}
\end{figure}
\begin{figure}[hbt!]
\begin{center}
\includegraphics [angle=0,scale=0.41]{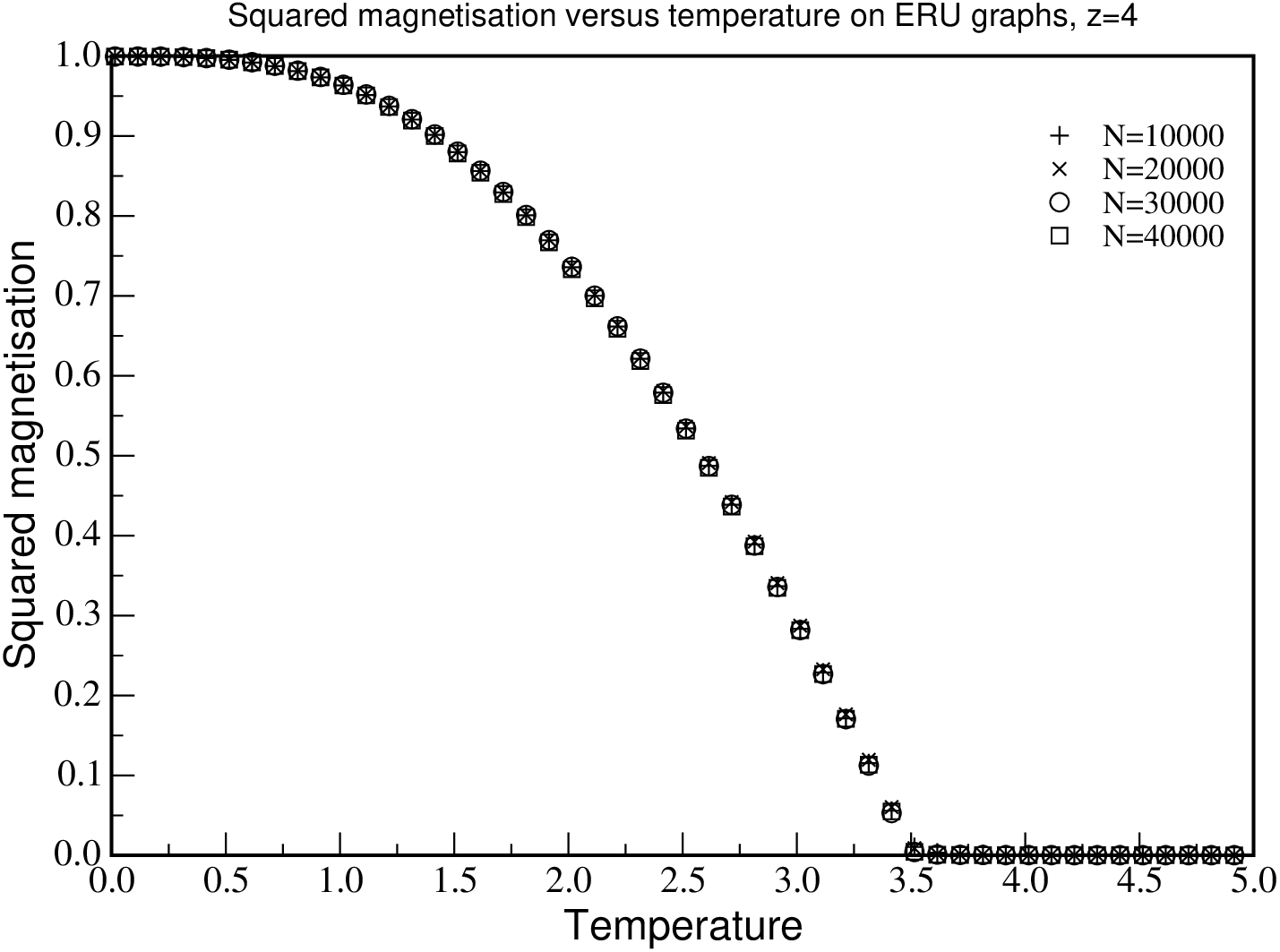}
\end{center}
 \end{figure}
 \begin{figure}[hbt!]
 \begin{center}
\includegraphics [angle=0,scale=0.41]{Lima_Sumuor2b.eps}
 \end{center}
\caption{Squared normalized magnetization versus temperatures, $S=1/2,\; p=4/N$,
 for different sizes $N$ of the {\it undirected} ER graph (top) and {\it directed} ER graph (bottom).}
\end{figure}
\bigskip
\bigskip
\begin{figure}[hbt]
\begin{center}
\includegraphics [angle=-90,scale=0.45]{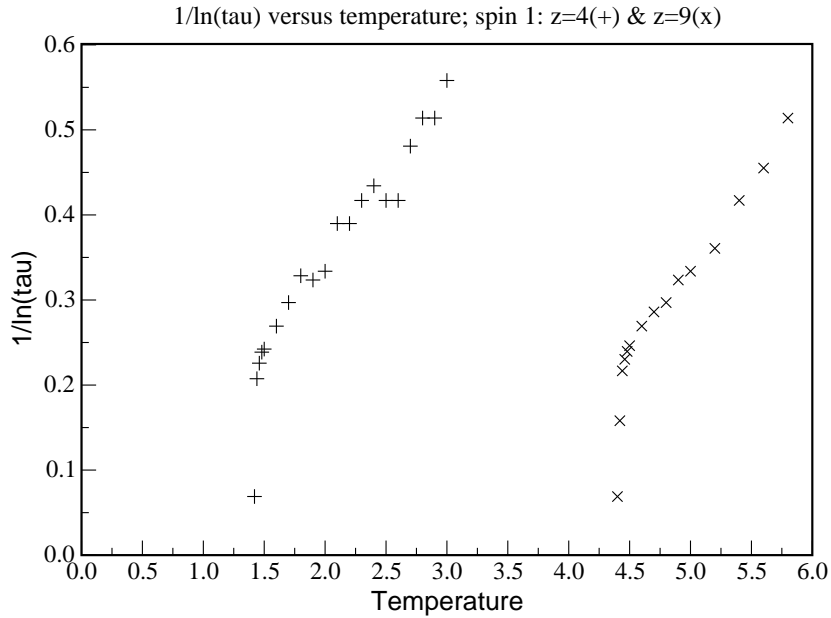}
\end{center}
\caption{Reciprocal logarithm of the relaxation times on {\it directed} ER networks versus $T$, with $S=1$ for different probabilities $p=z/N$ with $z=2$(+) and 9(x), $N=4,000,000$.}
\end{figure}
\begin{figure}[hbt!]
\begin{center}
\includegraphics [angle=0,scale=0.4]{Lima_Sumuor4a.eps}
\end{center}
\end{figure}
\begin{figure}[hbt!]
\begin{center}
\includegraphics [angle=0,scale=0.4]{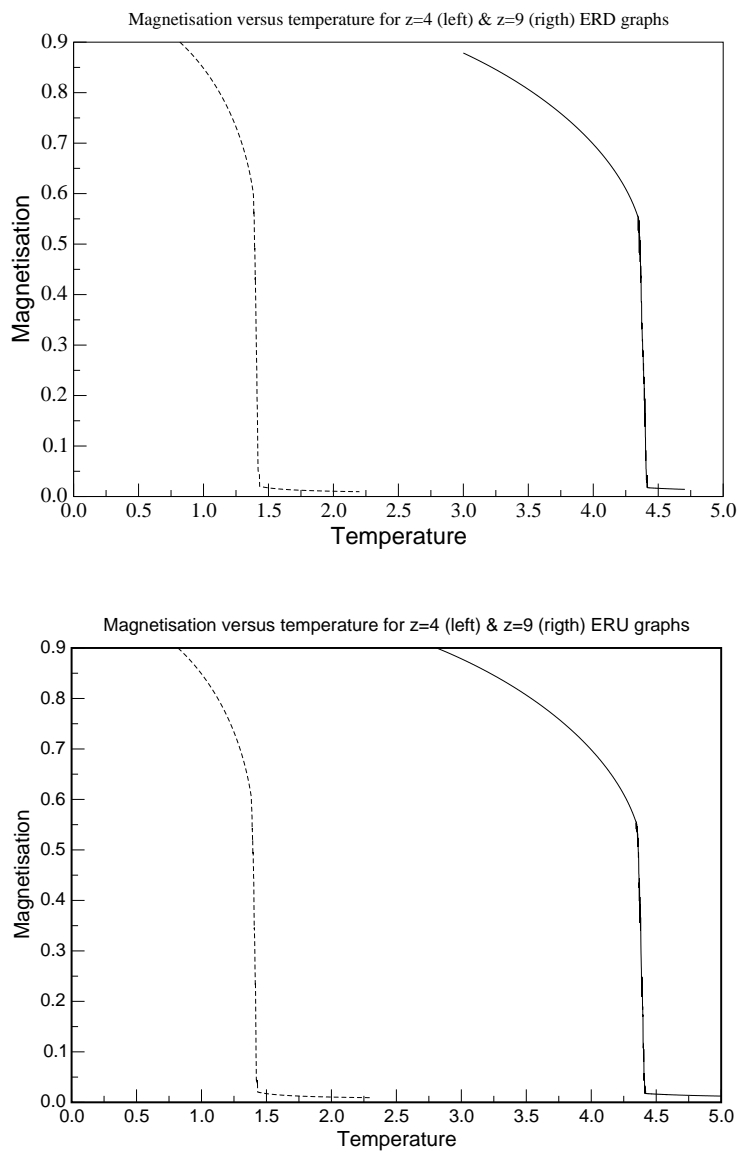}
\end{center}
\caption{ Magnetisation versus temperature for spin $S=1$ on directed (top) and undirected (bottom) ER graphs.}
\end{figure}
\begin{figure}[hbt]
\begin{center}
\includegraphics [angle=0,scale=0.45]{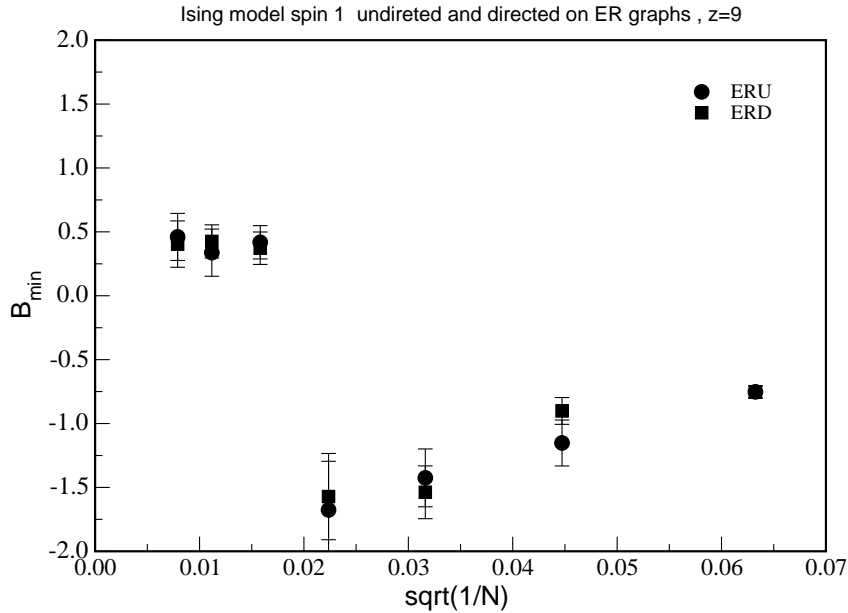}
\end{center}
\caption{Energetic Binder Cumulant $B_{\min}$ versus $1/N$ for $z=9$.}
\end{figure}

\begin{figure}[hbt]
\begin{center}
\includegraphics [angle=0,scale=0.45]{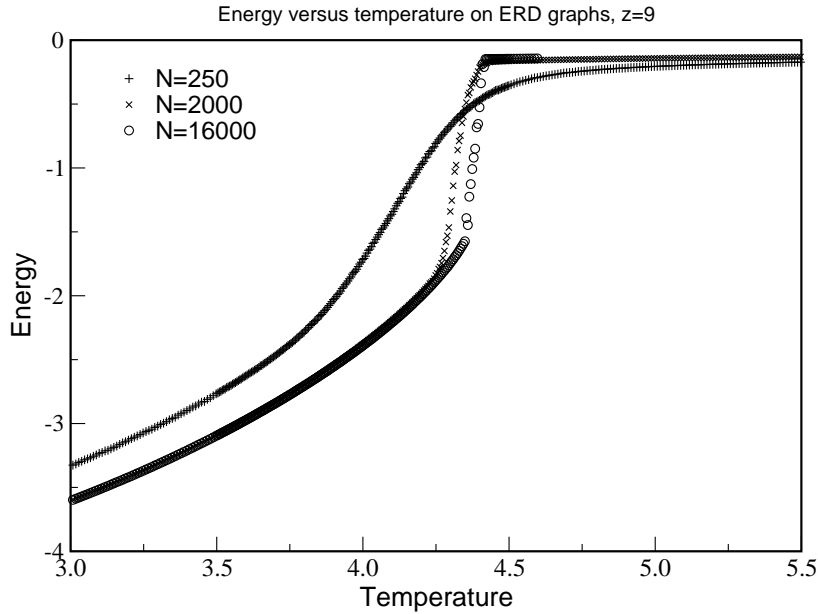}
\end{center}
\caption{Energy versus temperature on ERD graphs for $z=9$.}
\end{figure}

\bigskip
{\bf Model and Simulation: Ising model on ER graphs}

We consider the spin $S=1/2$ and 1 Ising models defined by a set of spin variables $S_i$
located on every site $i$, first of {\it directed} ER graphs, with $N$ spins taking the values $\pm 1$
and 0 for $S=1$, and $\pm 1$ for $S=1/2$, respectively.

The probability for spin $S_{i}$ to change its state in this {\it directed} 
network is
\begin{equation}
p_{i}= 1/[1+\exp(-2E_{i}/k_BT)],  \quad E_{i}=-J\sum_{k}S_{i}S_{k}
\end{equation}
and enters the heat bath algorithm; $k$ runs over all nearest neighbors of 
$S_{i}$. In this network, each new site
added to the network selects with connectivity $z$ already existing sites as neighbors
influencing it; the newly added spin does not influence these neighbors. 

To study the spin $1/2$ and $1$ Ising models we start with all spins up, a number of
spins equal to $2,000,000$ and $4,000,000$, and Monte Carlo step (MCS) time up to $200,000$ and $2,000,000$, respectively. In our
simulations, one MCS is accomplished after all  spins are updated, here, with heat bath
Monte Carlo algorithm. Then we vary the temperature and study nine samples. The
temperature is measured in units of the critical temperature of the square-lattice Ising
model. We determine the time $\tau$ after which the magnetization has flipped its sign
for the first time, and then take the median value of our nine samples. So we get
different values $\tau$ for different temperatures. 
To study the critical behavior of this Ising model (with spins 1/2 and 1) we
define the variable $m=\sum_{i=1}^{N}S_{i}/N$ as normalized magnetization.
The Ising model on directed BA networks has no phase transition and agrees with the modified Arrhenius law for relaxation time,$1/\ln(\tau) \propto T +...$, Lima
et al. \cite{lima2}.

\newpage
{\bf Results and Discussion}
\bigskip

{\bf Spin 1/2 Ising model} 

We take different probabilities for different number of nodes $N=2,000,000$  with different temperatures in Fig. 1. There we check the first time
after which the magnetization changes sign, take the median from nine samples,
and plot the reciprocal of the time for three probabilities $p=z/N$ ($z=1$, 2, 
and 3) in Fig. 1.
The  figure shows nicely the difference between probability $1/N$ (= percolation threshold) and larger probabilities. This  figure shows that there is a spontaneous magnetization at $p = 2/N$ for the left curve and at $p = 3/N$
for the right curve, but
no spontaneous magnetization at $p = 1/N$ which is the percolation threshold.
In Fig. 2  we show the dependence of the magnetization $M$  on the temperature, obtained
for directed and undirected ER graphs with $S=1/2$, we use only one probability equal $p = 4/N$, because it gives a clear answer compatible with the mean-field universality class, as expected
because of the infinite range of the symmetric interaction. For {\it undirected} ER graphs, if A is a neighbor of
B then, in contrast to the directed case, also B is a neighbor of A. From our simulation we see that the undirected
version has a spontaneous magnetization, to which the system relaxes similarly to
the standard Ising square lattice. Then we plot the square of normalized magnetization
versus temperature in Fig. 2. For $T$ below $T_c$ we have a spontaneous magnetization and above $T_c$ we do not have one as we see in Fig. 2 (part (a)). In equilibrium there is a Curie temperature. The squared magnetization vanishes at this $T_c \approx 3.5J/K_B$
linearly in temperature. This behavior corresponds, not unexpectedly, to a mean
field critical exponent. Unexpectedly, this same behavior occurs also for {\it directed} ER graphs (part (b)) that do not present an infinite range of the symmetric interaction as occurs with {\it undirected} ER graphs. The squared magnetization vanishes at this $T_c \approx 1.2J/K_B$. These results show that the behaviors of $S=1/2$ Ising model spin on ER graphs are similar,
whether these networks are { \it directed} or {\it undirected}.
\bigskip
\newpage
{\bf Spin 1 Ising model}

Fig. 3 is analogous to Fig. 1 except that now $S=1$ instead of 1/2 for $N=4,000,000$ up sites. 
In Fig. 4  we show magnetisation versus temperature on {\it directed} ER networks (part (a)) and also on {\it undirected} ER networks (part (b)) for different probabilities $p=z/N$ with $z=4$ (left) and $9$ (right) for system size $N=16,000$ sites. The shapes of these figures show qualitatively that they present evidence of first-order phase transition and also show that the behaviors of magnetisation versus temperature are identical for the same probabilities regardless of whether the networks are {\it directed} or {\it undirected}. 
In order to verify the order of the transition, we apply finite-size scaling (FSS) for $N=250$, $500$, $1,000$, $2,000$, $4,000$, $8,000$, and $16,000$ sites. Initially we search for the minima of the energetic fourth-order cumulant:
\begin{equation}
B = 1- \left[\frac{<e^4>}{3<e^2>^2}\right]_{av}
\end{equation}
It is known that this parameter takes a minimum value $B_{\min}$ at the effective transition temperature $T_{c}(N)$. One can show \cite{pri} that for a second-order transition $\lim_{N \to \infty} (2/3-B_{\min})=0$, even at $T_{c}$, while at a first-order transition the same limit is different from zero ($\neq 0$).
In Fig. 5 we plot the Binder minimum parameter $B_{\min}$ versus $1/N$ (eq. (2))
for $z=9$, and several system sizes. The Binder parameter goes to a value which is different from 2/3. This is a sufficient condition to characterize a first-order transition. The order of transition can be confirmed by plotting the values of energy  versus temperature, see Fig. 6, where we present a jump when system sizes increase. This behavior is evidence for a first-order phase transition for $z=9$, this same behavior occurs also for $z=4$.

\bigskip
{\bf Conclusion}

In conclusion, we have presented the Ising model for spins $S=1/2$ and $1$  on {\it
directed} ER  and {\it undirected} ER graphs, because our main objective in this paper
was to verify the existence or not of phase transitions and also the kind of phase transition. 

For spin $S=1/2$ Ising models, both {\it directed} or {\it undirected} ER graphs have a phase transition temperature below which a spontaneous magnetization exists, where  ER graphs have a spontaneous magnetization in the universality class of mean  field theory. For spin $S=1$ Ising models, on {\it directed} and {\it undirected} ER graphs the results are identical, i.e, are independent of the nature of the graphs studied here and have both a good evidence of a first-order phase transition different from spin $S=1/2$. Our results agree with the results of nonequilibrium model on directed and undirected ER graphs studied for  Pereira et al. \cite{fe} and Lima et al. \cite{lima3}.

The authors thank  D. Stauffer for many suggestions and fruitful discussions during the
development this work and also for the revision of this paper. We also acknowledge the
Brazilian agency CNPQ for  its financial support. This
work also was supported the system SGI Altix 1350 the computational park
CENAPAD.UNICAMP-USP, SP-BRAZIL.

\end{document}